\documentclass[twocolumn]{aastex631}
\usepackage{showyourwork}

\newcommand{\Msun}{\ensuremath{\rm{M}_{\odot}}\xspace}

\newcommand{\yr}{\ensuremath{\,\rm{yr}}\xspace}

\newcommand{\Gyr}{\ensuremath{\,\rm{Gyr}}\xspace}
\newcommand{\Mpc}{\ensuremath{\,\rm{Mpc}}\xspace}

\newcommand{\Mbheen}{\ensuremath{\,M_{\rm BH, 1}}\xspace}

\newcommand{\SFRDzZ}{\ensuremath{\mathcal{S}(Z,z)}\xspace} 
\newcommand{\SFRDz}{\ensuremath{\mathrm{SFRD}(z)}\xspace} 
\newcommand{\dPdZ}{\ensuremath{\mathrm{\frac{dP}{dZ}}(Z,z)}\xspace}
\newcommand{\dpdZ}{\ensuremath{\mathrm{dP/dZ}(Z,z)}\xspace}

\newcommand{\COMPAS}{{\tt COMPAS}\xspace}

\usepackage{xspace}
\usepackage{cancel}
\usepackage{amsmath}

\begin{document}

\title{The locations of features in the mass distribution of merging binary black holes are robust against uncertainties in the metallicity-dependent cosmic star formation history.}

\correspondingauthor{L.~van Son}
\email{lieke.van.son@cfa.harvard.edu}

\author[0000-0001-5484-4987]{L.~A.~C.~van~Son}
\affiliation{Center for Astrophysics $|$ Harvard $\&$ Smithsonian, 60 Garden St., Cambridge, MA 02138, USA}
\affiliation{Anton Pannekoek Institute of Astronomy, Science Park 904, University of Amsterdam, 1098XH Amsterdam, The Netherlands}
\affiliation{Max Planck Institute for Astrophysics, Karl-Schwarzschild-Str. 1, 85748 Garching, Germany}
\author[0000-0001-9336-2825]{S. E. de Mink}
\affiliation{Max Planck Institute for Astrophysics, Karl-Schwarzschild-Str. 1, 85748 Garching, Germany}
\affiliation{Anton Pannekoek Institute of Astronomy, Science Park 904, University of Amsterdam, 1098XH Amsterdam, The Netherlands}
\affiliation{Center for Astrophysics $|$ Harvard $\&$ Smithsonian, 60 Garden St., Cambridge, MA 02138, USA}
\author[0000-0002-8901-6994]{M. Chru{\'s}li{\'n}ska}
\affiliation{Max Planck Institute for Astrophysics, Karl-Schwarzschild-Str. 1, 85748 Garching, Germany}
\author[0000-0002-1590-8551]{C. Conroy}
\affiliation{Center for Astrophysics $|$ Harvard $\&$ Smithsonian, 60 Garden St., Cambridge, MA 02138, USA}
\author[0000-0003-3308-2420]{R. Pakmor}
\affiliation{Max Planck Institute for Astrophysics, Karl-Schwarzschild-Str. 1, 85748 Garching, Germany}
\author[0000-0001-6950-1629]{L. Hernquist}
\affiliation{Center for Astrophysics $|$ Harvard $\&$ Smithsonian, 60 Garden St., Cambridge, MA 02138, USA}

\begin{abstract}
New observational facilities are probing astrophysical transients such as stellar explosions and gravitational wave (GW) sources at ever increasing redshifts, while also revealing new features in source property distributions.
To interpret these observations, we need to compare them to predictions from stellar population models. Such models require the metallicity-dependent cosmic star formation history (\SFRDzZ) as an input. Large uncertainties remain in the shape and evolution of this function.
In this work, we propose a simple analytical function for \SFRDzZ. 
Variations of this function can be easily interpreted, because the parameters link to its shape in an intuitive way. 
We fit our analytical function to the star-forming gas of the cosmological TNG100 simulation and find that it is able to capture the main behaviour well. 
As an example application, we investigate the effect of systematic variations in the \SFRDzZ parameters on the predicted mass distribution of locally merging binary black holes (BBH). Our main findings are: I) the locations of features are remarkably robust against variations in the metallicity-dependent cosmic star formation history, and II) the low mass end is least affected by these variations. This is promising as it increases our chances to constrain the physics that governs the formation of these objects.
\end{abstract}

\section{Introduction \label{sec: intro}}
A myriad of astrophysical phenomena depend critically on the rate of star formation throughout the cosmic history of the Universe. Exotic transient phenomena, including (pulsational) pair-instability supernovae, long gamma-ray bursts and gravitational wave (GW) events appear to be especially sensitive to the metallicity at which star formation occurs at different epochs throughout the Universe \citep[e.g.,][]{Langer+2007,Fruchter+2006,LIGO2016_implications}.
Gravitational astronomy in particular has seen explosive growth in the number of detections in the past decade \citep[][]{GWTC1,GWTC2,GWTC3}, while theoretical predictions vary greatly due to uncertainties in the aforementioned metallicity of star formation \citep[e.g.,][]{Santoliquido+2021,Broekgaarden+2021b}. In order to correctly model and interpret these observations, it is thus fundamental to know the rate of star formation at different metallicities throughout cosmic history; i.e. the metallicity-dependent cosmic star formation history \citep[\SFRDzZ, see also the recent review by][]{chruslinska2022_review}. Throughout this work little $z$ refers to the redshift and $Z$ to the metallicity of star formation.

It is difficult to observationally constrain the shape of \SFRDzZ\ -- \citep[see e.g., ][for discussion of relevant observational caveats]{Chruslinska2019_obs,Boco+2021}. Even at low redshifts, the low metallicity part of the distribution is poorly constrained \citep{Chruslinska+2021}.
Nonetheless, several methods exist to estimate the metallicity-dependent cosmic star formation history. 

The first method is based on empirical scaling relations, linking galaxy properties like stellar mass $M_{\star}$, metallicity $Z$, and overall star-formation rate density \SFRDz, with the galaxy stellar mass function, GSMF \citep[see e.g.][]{Dominik+2013}. However, the applied methods to infer galaxy properties and subsequently scaling relations such as the MZ-relation differ greatly, which makes it difficult to interpret these results in a consistent way \citep[e.g.,][]{KewleyEllison2008,MaiolinoMannucci2019,Cresci+2019}. Moreover, observations are generally incomplete at high redshifts and low galaxy luminosity \citep[e.g.,][]{Chruslinska+2021}.

One can also directly extract the metallicity-dependent cosmic star formation history from cosmological simulations \citep[e.g.][]{Mapelli2017, Briel+2022_rates}. However, these simulations currently lack the resolution to resolve the lowest mass galaxies, and their variations in \SFRDzZ span a smaller range than those observed in observationally-based models \citep{Pakmor+2022}.

Alternatively, one can combine analytical models for the observed overall star-formation rate density, \SFRDz, like those from \cite{MadauDickinson2014} or \cite{Madau+2017}, and convolve this with an assumed function for the shape of the cosmic metallicity density distribution, such as was was done in e.g., \cite{LangerNorman2006} and the phenomenological model in \cite{Neijssel+2019}.

In this work we follow the latter approach and propose a flexible analytical model for \SFRDzZ that can be fit to the output of both cosmological simulations, and observational data constraints where available. 
In contrast to earlier work, we adopt a skewed-lognormal distribution of metallicities that can capture the asymmetry in the low and high metallicity tails. 

The purpose of this proposed form is twofold.  
First of all, the form we propose allows for an intuitive interpretation of the free parameters. This allows us to get better insight of the impact of changes in these parameters on the inferred ranges of astrophysical transients (as we demonstrate in Section \ref{sec: mass dists} using GW predictions as an example). By adopting an analytical, parametrized form for \SFRDzZ, the large uncertainties can be systematically explored.
Secondly, both the large complications in observational constraints, and the many uncertainties in cosmological simulations call for a generalised form of \SFRDzZ that can be easily updated when new information becomes available. 
In particular, the advent of observations with the James Webb Space Telescope promises a new era of high-redshift
metallicity studies of previously unexplored regimes \citep[e.g.,][]{Sanders+2022}. We hope that this form will facilitate the flexibility needed to keep up with observations. 
The model described in this work is incorporated in the publicly available `Cosmic Integration' suite of the \COMPAS code.\footnote{\url{https://github.com/TeamCOMPAS/COMPAS/tree/dev/utils/CosmicIntegration}}

We describe our model for \SFRDzZ in Section \ref{sec: model for sfrd(zZ)}.
We fit our model to the star-forming gas in the Illustris TNG100 simulation in Section \ref{sec: fit against tng}, and demonstrate an example application of our model by systematically varying the parameters that determine the shape of \SFRDzZ and investigate their impact on the local distribution of merging BBH masses in Section \ref{sec: mass dists}.
We summarise our findings in Section \ref{sec: summary}.

Throughout this work, we adopt a universal Kroupa initial mass function \citep{Kroupa2001} with the mass
limits $0.01 - 200\Msun$ and a flat $\Lambda$CDM cosmology with $\Omega_{\rm{M}}=0.31$, $\Omega_{\rm{\Lambda}}=0.69$ and $H_0=67.7\rm{km\,s^{-1}\,Mpc^{-1}}$ \citep{Planck18_VI}.

\section{A convenient analytic expression for the metallicity-dependent cosmic star formation history \label{sec: model for sfrd(zZ)} }
We write the metallicity-dependent cosmic star formation history as

\begin{equation}
\label{eq: total sfrd}
\boxed{
        \SFRDzZ = \SFRDz \times \dPdZ
        }
\end{equation}
(similar to e.g., \citealt{LangerNorman2006}).
The first term is the star formation rate density, \SFRDz, that is the amount of mass formed in stars per unit time and per unit comoving volume at each redshift, $z$. The second term, \dpdZ, is a probability density distribution that expresses what fraction of star formation occurs at which metallicity, $Z$, at each redshift. 
 
\subsection{The cosmic metallicity density distribution}
For the probability distribution of metallicities we draw inspiration from the approach by e.g., \cite{Neijssel+2019} who used a log-normal distribution for their phenomenological model. Unfortunately, a simple log-normal distribution cannot capture the asymmetry that we see in the cosmological simulations, which show an extended tail in  $\log_{10} Z$ towards low metallicity, combined with a very limited tail towards higher metallicity. To capture this behaviour we adopt a skewed-log-normal distribution instead. This is an extension of the normal distribution that introduces an additional shape parameter, $\alpha$, that regulates the skewness \citep[first introduced by][]{Ohagan+1976}. 

The skewed-log-normal distribution of metallicities is defined as:
\begin{equation}
\begin{aligned}
\label{eq: pure log skew}
\mathrm{\frac{dP}{dZ}}(Z, z) &= \frac{1}{Z} \times \frac{\mathrm{dP(}Z,z) }{{\rm d}\ln Z}  \\
&= \frac{1}{Z} \times \frac{2}{\omega}
    \underbrace { \phi \left(\frac{\ln Z - \xi}{\omega}\right)
                 }_{(a)}
    \underbrace {
                \Phi\left(\alpha \frac{\ln Z - \xi}{\omega} \right)
                }_{(b)},
\end{aligned}
\end{equation}

\noindent where (a) is the standard log-normal distribution, $\phi$,
\begin{equation}
\label{eq: log normal and CDF}
 \phi \left(\frac{\ln Z - \xi}{\omega}\right) \equiv 
    \frac{1}{\sqrt{2 \pi}} 
    \exp{
         \left\{
            -\frac{1}{2} \left(\frac{\ln Z - \xi}{\omega}\right)^2
        \right\}
        }
    \end{equation}
and (b) is the new term that allows for asymmetry, which is equal to the cumulative of the log-normal distribution, $\Phi$,
    \begin{equation}
    \begin{array}{cc}
 \Phi\left(\alpha \frac{\ln Z - \xi}{\omega} \right) &\equiv 
    \frac{1}{2} 
    \left[ 
        1 + {\rm erf}
            \left\{
                \alpha \frac{\ln Z - \xi}{\omega \sqrt{2}}
            \right\} 
    \right] . \\
    \end{array}
\end{equation}

\noindent This introduces three parameters, $\alpha, \omega$ and $\xi$, each of which may depend on redshift. The first parameter, $\alpha$, is known as the ``shape''. It affects the skewness of the distribution and thus allows for asymmetries between metallicities that are higher and lower than the mean.  The symmetric log-normal distribution is recovered for $\alpha=0$. The second parameter, $\omega$  is known as the ``scale''. It provides a measure of the spread in metallicities at each redshift.   Finally, $\xi$, is known as the ``location'', because this parameter plays a role in setting the mean of the distribution at each redshift.

\paragraph{The location and the mean of the metallicity distribution}
To obtain a useful expression for the redshift dependence of the ``location'' $\xi(z)$ we first express the expectation value or mean metallicity at a given redshift
\begin{equation}
 \langle  Z \rangle 
 = 2 \exp
        \left(\xi +  \frac{\omega^2}{2} \right)
         \Phi\left(\beta\, \omega\right)
 \label{eqn:Zmean}
\end{equation}
where $\beta$ is 
\begin{equation}
\label{eq: beta}
\beta = \frac{\alpha}{\sqrt{1 + \alpha^2} }.
\end{equation}
(For a more extended derivation of the moments of the skewed-log-normal, see e.g., \cite{WANG201995}.)

For the evolution of the mean metallicity with redshift we follow \cite{LangerNorman2006} and the phenomenological model from \cite{Neijssel+2019} in assuming that the mean of the probability density function of metallicities evolves with redshift as:
\begin{equation}
\label{eq: mean Z}
    \langle Z \rangle \equiv \mu(z) = \mu_0 \cdot 10^{\mu_z \cdot z},
\end{equation}
where $\mu_0
$ is the mean metallicity at redshift 0, and $\mu_z
$ determines redshift evolution of the location. Equating this to Equation~\ref{eqn:Zmean}, we get an expression for $\xi(z)$,

\begin{equation}
\label{eq mu z}
    \xi(z) = \ln\left(\frac{  \mu_0 \cdot 10^{\mu_z \cdot z} }{2\, \Phi(\beta\, \omega)}  \right) - \frac{\omega^2}{2}.
\end{equation}

\paragraph{The scale (and variance) of the metallicity distribution}

We will also allow the ``scale'' $\omega$ to evolve with redshift in a similar manner, 
\begin{equation}
\label{eq: omega z}
    \omega(z) = \omega_0 \cdot 10^{\omega_z \cdot z}.
\end{equation}
where $\omega_0$ is
the width of the metallicity distribution at $z=0$, and $\omega_z$
the redshift evolution of the scale.

Note that the width, $w(z)$ is not the same as the variance. The variance, $\sigma^2(z)$, can be expressed as

\begin{equation}
    \sigma^2(z) = \omega^2(z) \left( 1 - \frac{2\beta^2}{\pi} \right)
\end{equation}

\paragraph{Asymmetry of the metallicity distribution: $\alpha$}
The skewness $\alpha$ could in principle also be allowed to evolve with redshift (e.g., $\alpha (z) = \alpha(z=0) 10^{\alpha_z \cdot z}$). 
However, we find no significant improvement over the simpler assumption where alpha is kept constant. 
Note that the redshift evolution of the `scale' (eq. \ref{eq: omega z}), already captures similar behaviour in our current formalism. We therefore adopt $ \alpha = \alpha(z=0)$ and $\alpha_z = 0$.

In summary, Equation~\ref{eq: pure log skew} becomes:
\begin{equation}
\label{eq: z log skew}
\boxed{
    \dPdZ = \frac{2}{\omega(z) Z} \times \phi \left(\frac{\ln Z - \xi(z)}{\omega(z)}\right) \Phi\left(\alpha \frac{\ln Z - \xi(z)}{\omega(z)} \right)
    } \ , 
\end{equation}

\noindent where $\xi(z)$ and $\omega(z)$ are defined in Equations~\ref{eq mu z} and \ref{eq: omega z} respectively and we have assumed $\alpha$ to be constant.

\subsection{The overall cosmic star formation rate density}
For the star formation rate density, we assume the analytical form proposed by \cite{MadauDickinson2014},
\begin{eqnarray}
\label{eq: sfr1}
\boxed{
    \SFRDz  = 
    \frac{d^2 M_{\rm SFR}}{dt dV_c} (z)= 
    a \frac{\left(1 + z\right)^b}{1 + \left[ (1 + z)/c \right]^d} 
    }\,
\end{eqnarray}
in units of $\left[ \Msun \,yr^{-1} \,cMpc^{-3} \right]$. This introduces four parameters: $a$ which sets the overal normalisation and which has the same units as \SFRDz and $b,c$ and $d$ which are unitless and which govern the shape of the overal cosmic star formation rate density with redshift. \\

Lastly, we combine equations \ref{eq: z log skew} and \ref{eq: sfr1} to form a full metallicity specific star formation rate density as described in equation \ref{eq: total sfrd}.

\section{Fit against Cosmological simulation \label{sec: fit against tng}}

We fit our new functional form of \SFRDzZ as defined by equations \ref{eq: total sfrd}, \ref{eq: z log skew} and \ref{eq: sfr1} to the IllustrisTNG cosmological simulations. 
We simultaneously fit for the following nine free parameters $\alpha, \mu_0, \mu_z, \omega_0, \omega_z$, which govern the metallicity dependence and $a,b, c$ and $d$, which set the overall star-formation rate density.
Below we briefly discuss the IllustrisTNG simulations, and elaborate on our fitting procedure.

\subsection{IllustrisTNG Cosmological simulations}
Although here, we only fit our model to the TNG100 simulation, our prescription can be easily be used to fit other simulated or observational data of the metallicity-dependent cosmic star formation history\footnote{We provide a Jupyter notebook to facilitate this fit here: \url{https://github.com/LiekeVanSon/SFRD_fit/blob/main/src/scripts/Notebooks/Fit_model_to_sfrdzZ.ipynb} } .

The IllustrisTNG-project (or TNG in short) considers galaxy formation and evolution through large-scale cosmological hydrodynamical simulations \citep[][]{FirstResTNG_Springel2018,FirstResTNG_Marinacci2018, FirstResTNG_Nelson2018,FirstResTNG_Pillepich2018, FirstResTNG_Naiman2018, Nelson2019a, Pillepich2019}.
Such simulations provide the tools to study parts of the Universe that are not easily accessible by observations. In particular of interest for this work, they simulate the high redshift enrichment of galaxies and the tail of low metallicity star formation at low redshift.

The models implemented in the publicly available TNG simulations \citep{Nelson2019b}\footnote{ \url{https://www.tng-project.org/}} have lead to many successes. 
These models where calibrated at the resolution of the TNG100 simulation, hence TNG100 is expected to provide the best overall agreement to global properties (like the star formation rate density). This is why we adopt the TNG100 simulation as our fiducial simulation.
For a more extended discussion focused on the processes that govern the creation, distribution and mixing of metals in in the TNG simulations, we refer to \cite{Pakmor+2022}. In short, star formation in the TNG simulations is calibrated against the Kennicutt–Schmidt relation \citep[][]{Schmidt1959,Kennicutt1989}, using an effective equation of state \citep{SpringelHernquist2003}. The stellar metallicity yields are an updated version of the original Illustris simulations as described in \cite{Pillepich2018}. Star particles deposit metals into the gas through type Ia and type II supernovae, as well as through asymptotic giant branch stars. 
The TNG simulations have been shown to match observational constraints on the mass-metallicity relation of galaxies up to $z = 2$ \citep{Torrey+2019}, as well as iron abundances \citep{FirstResTNG_Naiman2018}, metallicity gradients within galaxies at low redshift \citep{Hemler+2021}, and the reduction of star formation in the centers of star-forming galaxies \citep{Nelson2021}. 
Several studies have used the TNG simulations to make predictions for astronomical transient sources \citep[e.g.][]{Briel+2022_rates,Bavera+2022,vanson+2022}. 
Out of the four \SFRDzZ variations explored, \cite{Briel+2022_rates} find that TNG provides one of the best agreements between observed and predicted cosmic rates for electromagnetic and gravitational-wave transients, when combined with their fiducial binary population synthesis model. 

On the other hand, large uncertainties and crude approximations remain in all contemporary cosmological simulations, thus also in the TNG simulations. 
Generally, some of the chemical evolution of galaxies in cosmological simulations is unresolved, and thus depends strongly on the implemented `sub-grid physics'.
A known uncertainty is that dust is not included in the TNG simulations, which could mean that metallicity of the star-forming gas is overestimated. 
Feedback from active galactic nuclei is not well understood theoretically and is described in an approximate manner \citep{Springel2005, Weinberger2017}.  Furthermore, all stellar winds mass loss from massive stars, binary interactions and their ionising effects are ignored \citep[e.g.][]{Dray+2003,Smith2014,Gotberg+2020,DoughtyFinlator2021,Farmer2021_carbonfootprint,Goswami+2022}.
Moreover, the uniform ionising UV background is turned on abruptly at $z=6$. This crucially impacts the amount of low metallicity star formation at high redshift as it allows small galaxies to produce more stars than what would be expected for a gradually increasing UV background that reaches full strength at $z=6$.
All these uncertainties underline the need for a flexible approximation of the \SFRDzZ, that can be easily updated when cosmological models and sub-grid physics are updated.

\subsection{Choices and binning of the data}
We fit equation \ref{eq: total sfrd} to the metallicity-dependent star formation rate of the star-forming gas in the TNG100 simulation. For this we use a binned version of the TNG data $\SFRDzZ_{\rm sim}$. We consider metallicities between $\log_{10} Z= -5$ to $\log_{10} Z= 0$ in 30 bins, where we use $Z_i$ to refer to the logarithmic centres of the bins. We ignore star formation in metallicities $\log_{10} Z \le -5$ as this accounts for less than 1\% of the total cosmic star formation rate in these simulations.
We consider bins in redshifts between $z=0$ and $z=10$, with a step size of $dz=0.05$, where $z_j$ refers to the centres of the bins. 

\subsection{Optimisation function}
To find a solution we use a method based on the sum of the quadratic differences between the simulations and our fit function. Using a vanilla $\chi$-squared approach does not serve our purposes very well as it does a poor job in fitting regions where the star formation is very low.  Using a $\chi$-squared approach on the logarithm of the function instead places far too much weight on trying to fit the star formation rate in regions where the rate is very low or not even significant.  After experimenting, we find that the following approach gives us satisfactory results. 

We first consider a given redshift $z_j$.  For this redshift we compute the sum of the squared residuals between the cosmological simulation and our fit. This is effectively the square of the $l^2$-norm:
\begin{equation}
\label{eq: chisquare}
    \chi^2 (z_j) \equiv \sum_{Z_i} \left( 
        \mathcal{S}(Z_i,z_j)_{\rm sim} - 
        \mathcal{S}(Z_i,z_j)_{\rm fit}\right)^2 .
\end{equation}
Here, the variable $Z_i$ runs over all metallicity bins.
We are particularly interested in properly fitting the low metallicity star formation at high redshifts. At high redshifts, the overall star-formation rate density is generally lower. To ensure that our fitting procedure gives sufficient weight to the behaviour at all redshifts, we introduce a penalisation factor to somewhat reduce the contribution of redshifts where the peak of cosmic star formation occurs, while increasing the weight at redshifts where the overall star-formation rate density is lower.  To achieve this we divide $\chi^2 (z_j)$ by the star formation $\sum_{Z_i} \mathcal{S}(Z_i,z_j)$ per redshift bin before adding the contribution of all redshifts.  Our final expression for the cost function reads

\begin{equation}
\label{eq: cost function}
    \chi  = \sum_{z_j} \frac{ \chi^2 (z_j) } 
        {\sum_{Z_i} \mathcal{S}(Z_i,z_j)}
\end{equation}

To minimize this cost funciton, we use \texttt{scipy.optimize.minimize} from SciPy v1.6.3 which implements the quasi-Newton method of Broyden, Fletcher, Goldfarb, and Shanno \citep[BFGS,][]{NocedalWright_numerical_optimization}. 

\subsection{Resulting \SFRDzZ}

\begin{figure*}
\centering
\script{FitComparison_3panelPlot.py}
\includegraphics[width=0.9\textwidth]{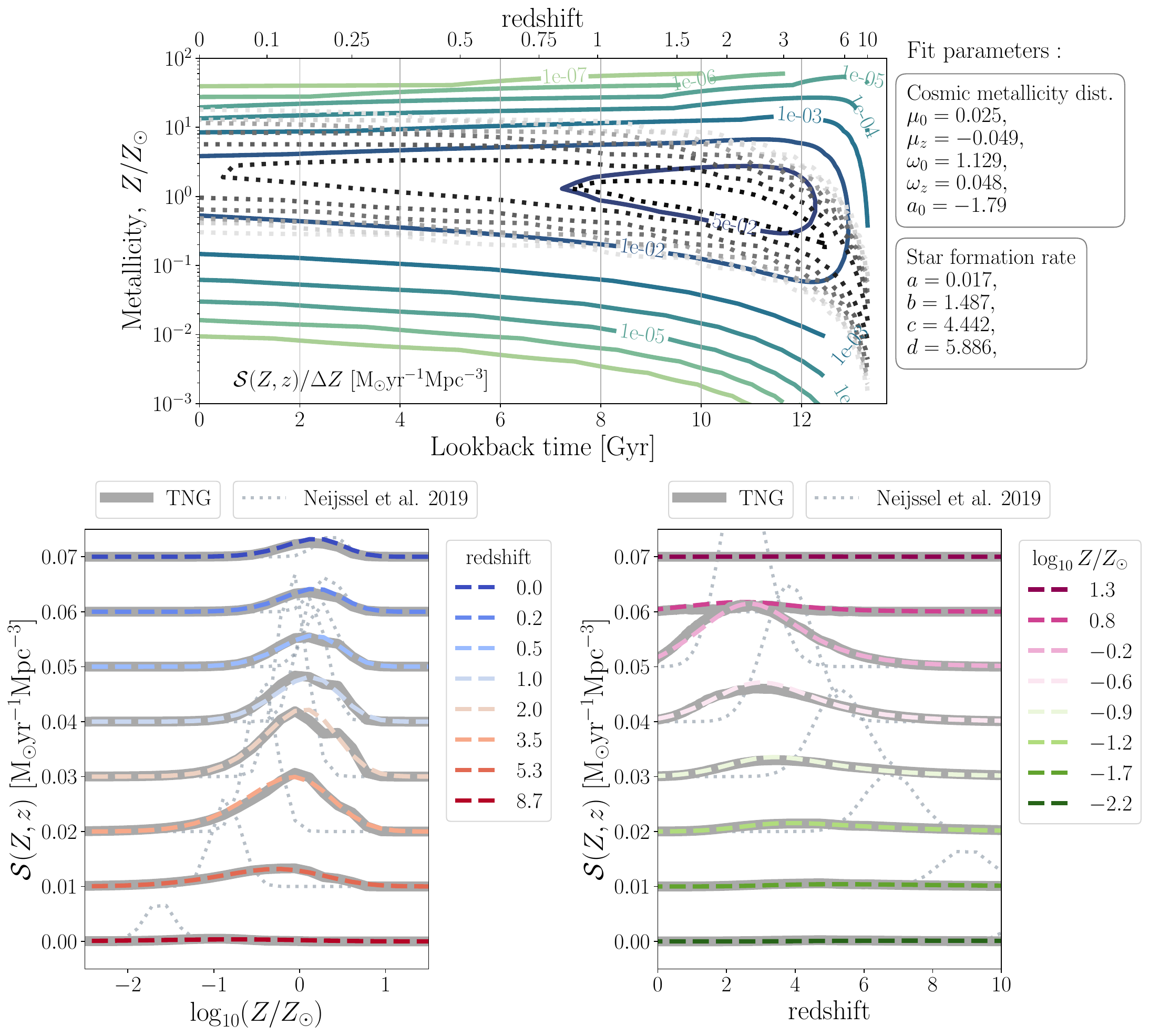}
\caption{Our fiducial \SFRDzZ model, adopting the best fitting parameters (listed on the top right) to fit the TNG100 simulations. The top panel shows the full two dimensional \SFRDzZ linear in time. Contours range from $10^{-7} - 10^{-2} \Msun \yr^{-1}\Mpc^{-3}$. 
The bottom left (right) panel shows slices of the distribution in redshift (metallicity). Each slice is displaced by 0.01$\Msun \yr^{-1}\Mpc^{-3}$ (note the linear scale of \SFRDzZ in the bottom panel). We show the TNG100 simulation data with thick gray lines. 
For comparison, we also show the phenomenological model from \protect\cite{Neijssel+2019} in all panels with grey dotted lines. The bottom panels show that our analytical model adequately captures the shape of the \SFRDzZ from TNG100.
 \label{fig: fit SFRD}}
\end{figure*}

Our best fitting parameters are listed in Table \ref{tab: fit params}. With these fit parameters, $\chi^2(z_j)$ is smaller than $2\cdot 10^{-4}$ at any given redshift. 
To evaluate our fit, we show the absolute residuals and relative errors in Appendix \ref{app: residuals}. 
We will refer to the \SFRDzZ with the parameters listed in Table \ref{tab: fit params} as our fiducial model. 

\begin{deluxetable*}{clr | cl}
\label{tab: fit params}
\tablecaption{Best fitting parameters for our \SFRDzZ fit to TNG100 data.}
\tablehead{\colhead{dP/dZ } & \colhead{description} & \colhead{best fit} &  \colhead{SFRD(z)} & \colhead{best fit} \\
 \colhead{ } & \colhead{} & \colhead{}  & \colhead{$\Msun\yr^{-1}\Mpc^{-3}$} & \colhead{} } 
\startdata
$\mu_0$    & mean metallicity at $z=0$    & $0.025  \pm 0.036$ &  $a$ &  $0.02 \pm 0.072$ \\
$\mu_z$    & $z$-dependence of the mean   & $-0.049 \pm 0.006$ &  $b$ &  $1.48 \pm 0.002$\\
$\alpha$   & shape (skewness)             & $-1.778 \pm 0.002$ &  $c$ &  $4.44 \pm 0.001$\\
$\omega_0$ & scale at $z=0$               & $1.122  \pm 0.001$ &  $d$ &  $5.90 \pm 0.002$\\
$\omega_z$ & $z$-dependence of the scale  & $0.049  \pm 0.009$ &   \\
\enddata
\end{deluxetable*}

In Figure \ref{fig: fit SFRD} we show our fiducial  model at different redshifts and metallicities. We also show the overall star-formation rate density \SFRDz in Figure \ref{fig: SFR(z)}.
In general, our analytical model captures the metallicity-dependent cosmic star formation history in the TNG100 simulations well (bottom panels of Figure \ref{fig: fit SFRD}). 
The skewed-log normal metallicity distribution is able to reproduce the overall behaviour that is observed in TNG100 \citep[bottom left panel, but cf. ][for an in-depth discussion of low metallicity star formation in the TNG50 simulation]{Pakmor+2022}.
Only minor features like the additional bump just above $\log_{10}(Z) = -2$ at redshift 2 are missed. 
However, for our purposes, it is more important to prioritise fitting the large scale trends, while we are not so interested in smaller scale fluctuations.

Adopting a skewed-lognormal metallicity distribution allows for a tail of low metallicity star formation out to low redshifts. To emphasise the difference between a skewed-lognormal and a symmetric lognormal distribution, we show the phenomenological model from \cite{Neijssel+2019} in dotted grey. Their model falls within the family of functions that is encompassed by our model described in Section \ref{sec: model for sfrd(zZ)}, but we note that their model is distinctly different.\footnote{The phenomenological model from \cite{Neijssel+2019} is recovered by adopting $\mu_0= 0.035$, $\mu_z=-0.23$, $\omega_0=0.39 $, $\omega_z = 0$, $\alpha = 0$, $a=0.01$, $b=2.77$, $c=2.9$ and $d=4.7$. }

Although our model preforms well at reproducing the large scale trends seen in TNG, we acknowledge that more complex features as suggested by some observational studies could be missed. One example is that the \SFRDz shape we adopt from \cite{MadauDickinson2014} does not account for starburst galaxies \citep[see discussion in][]{Chruslinska+2021}.
Moreover, our model cannot capture inflection points in the mean metallicity, because we assume both $\mu_0$ and $\mu_z$ are constants with redshift (equation \ref{eq: mean Z}). Contrarily, \cite{Chruslinska2019_obs} find an upturn in the amount of low metallicity star formation above $z=4$ if the power law of the GSMF is allowed to evolve with redshift.
Hence, although our model is more broadly applicable than previous models, in it's current form, it does not capture the complete range of observationally-allowed variations. 
Incorporating more complex functional forms for our the mean metallicity could possibly capture such behaviour, but this analysis is beyond the scope of this paper.

\section{Application: systematic variations of \SFRDzZ and the effect on the mass distribution of merging BBHs \label{sec: mass dists}}

We will now demonstrate the application of our analytical model by systematically varying the parameters in our fiducial \SFRDzZ model, and investigate their effect on the local mass distribution of BBH mergers originating from isolated binaries.

We use the publicly available rapid binary population synthesis simulations presented in \cite{vanson2022_lowMbh}.\footnote{Available for download at \url{https://zenodo.org/record/7612755},  see also the Software and Data section in the acknowledgements}  
These simulations were run using version v02.26.03 of the open source \COMPAS suite \citep{COMPAS_method}\footnote{\url{https://github.com/TeamCOMPAS/COMPAS}}. \COMPAS is based on algorithms that model the evolution of massive binary stars following \citet{Hurley+2000, Hurley+2002} using detailed evolutionary models by \citet{Pols+1998}. In particular, we use the simulations behind Figure 1 from \cite{vanson2022_lowMbh}, and we refer the reader to their methods section for a detailed description of the adopted physics parameters and assumptions. \footnote{ We note that the rate in \cite{vanson2022_lowMbh} is slightly higher than the fiducial rate presented in Figure \ref{fig: stable mass dists} in this work. This difference is caused by the use of rounded parameter values of \SFRDzZ in \cite{vanson2022_lowMbh}.}  
Metallicities of each binary system were sampled from a smooth probability distribution to avoid artificial peaks in the BH mass distribution \citep[e.g.][]{Dominik2015,Kummer_thesis}. 
These simulations provide us with an estimate of the yield of BBH mergers per unit of star-forming mass and metallicity. 

We combine the aforementioned yield with variations of the fiducial \SFRDzZ model described in this work. By integrating over cosmic history, we obtain the local merger rates of BBH systems, which allow us to construct the distribution of source properties at every redshift. We use the cosmic integration scheme that is part of the publicly available \COMPAS suite, which includes the \SFRDzZ model described in this work. The details of this framework are described in \cite{Neijssel+2019}, but also in \cite{vanson+2022}, where more similar settings to this work are used.

\subsection{Determining reasonable variations of \SFRDzZ \label{ss: reasonable var}}
We consider variations in both the shape of the cosmic metallicity density distribution \dpdZ, and the shape of the overall star-formation rate density, \SFRDz. To determine the range that is reasonably allowed by observations, we compare our variations to the observation-based \SFRDzZ models described in \cite{Chruslinska+2021}.
An overview of the explored variations is shown in Table \ref{tab: fit variations}. Below we explain how we arrive at these values.  

\begin{deluxetable}{cccc}
\label{tab: fit variations}
\tablewidth{0.5\textwidth} 
\tablecaption{Variations on \SFRDzZ. 
For every variation, we either swap the value of an individual \dpdZ parameter, or exchange the set of four \SFRDz parameters, and replace them by the the min/max values listed here. All other parameters are kept fixed at their fiducial value. }
\tablehead{\colhead{} & \colhead{min} & \colhead{fiducial} & \colhead{max} } 
\startdata
\dpdZ &       &  &  \\
$\mu_0$          &   0.007   & $0.025$  &  0.035  \\
$\mu_z$          &   0.0   & $-0.049$ &   -0.5 \\
$\alpha$         &   -6.0   & $-1.778$ &  0.0  \\
$\omega_0$       &   0.7   & $1.125$  &  2.0  \\
$\omega_z$       &   0.0   & $0.048$  &  0.1  \\ \hline
\SFRDz &       &  &  \\
($a$,$b$ ... &  ($0.01$, $2.60$   & ($0.02$, $1.48$ &  ($0.03$, $2.6$ \\
... $c$,$d$) &  $3.20$, $6.20$)   & $4.44$, $5.90$) &   $3.3$, $5.9$) \\
\enddata
\end{deluxetable}

For the cosmic metallicity density distribution, we vary every parameter that determines the shape of \dpdZ independently (three left-most columns of Table \ref{tab: fit params}, and top of Table \ref{tab: fit variations} ), while keeping all other parameters fixed at their fiducial value.
For each variation, we inspect the fraction of stellar mass that is formed at low-metallicity ($Z<0.1 Z_{\odot}$) versus the fraction of stellar mass that is formed at high-metallicity ($Z> Z_{\odot}$), for all star formation that occurred below a certain threshold redshift.
We compare this to the models from \cite{Chruslinska+2021} in Figure \ref{fig: low high Z fraction} in Appendix \ref{app: reasonable var}.
We have chosen our variations such that they span a reasonable range of cosmic metallicity density distributions as allowed by observation-based and cosmological simulations-based models.
We use the models \texttt{214-f14SB-BiC\_FMR270\_FOH\_z\_dM.dat}, and \texttt{302-f14SB-Boco\_FMR270\_FOH\_z\_dM.dat} from \cite{Chruslinska+2021}\footnote{These models including a detailed description of their contents  are publicly available at \url{https://ftp.science.ru.nl/astro/mchruslinska/Chruslinska_et_al_2021/} } as a representation of a very low and high metallicity star formation realisation respectively. These models are the low and high metallicity extreme under their fiducial SFR–metallicity correlation, and so we will refer to them as \texttt{Chr21\_lowZ} and \texttt{Chr21\_highZ} respectively from hereon. The difference between these models lies in the assumptions in the underlying empirical galaxy relations. In general, low-mass galaxies contribute to low-metallicity star formation and shift the peak of \SFRDzZ to lower metallicities. 
\texttt{Chr21\_lowZ} is characterised by a star formation–galaxy mass relation that is flat at high galaxy masses (reducing the star formation rate for the highest-mass galaxies), a galaxy stellar mass function that evolves with redshift (predicting an increasing number density of low-mass galaxies), 
and a local galaxy mass-metallicity relation as in \cite{PettiniPagel2004}. 
This model further approximates the contribution of starburst galaxies following \cite{Bisigello+2018} and \cite{Caputi+2017}. 
Assuming that starburst galaxies follow the empirical fundamental metallicity relation (leading to anti-correlation between the SFR and metallicity), their inclusion tends to shift the peak of \SFRDzZ to
lower metallicities and broadens the low-metallicity part of the distribution.

On the other hand, \texttt{Chr21\_highZ} assumes the star formation–galaxy mass relation does not flatten towards higher galaxy masses, a galaxy stellar mass function where the slope for the low-mass end is constant over redshift, and a local galaxy mass-metallicity relation following \cite{KobulnickyKewley2004}. Lastly, this model adopts the starburst prescription from \cite{Boco+2021}, which produces results that are similar to models without starburst galaxies. 

For every variation of our model, we inspect both the full \SFRDzZ and slices at redshifts $z = 0, 0.5, 3.0$ and $6$ by eye. At each slice we compare our model variation to \texttt{Chr21\_lowZ} and \texttt{Chr21\_highZ}, and ensure that none of our variations significantly exceeds these extremes in \SFRDzZ. This also serves as a sanity check for the overall star-formation rate density.

\begin{figure*}
\centering
\script{SFR_z.py}
\includegraphics[width=0.47\textwidth]{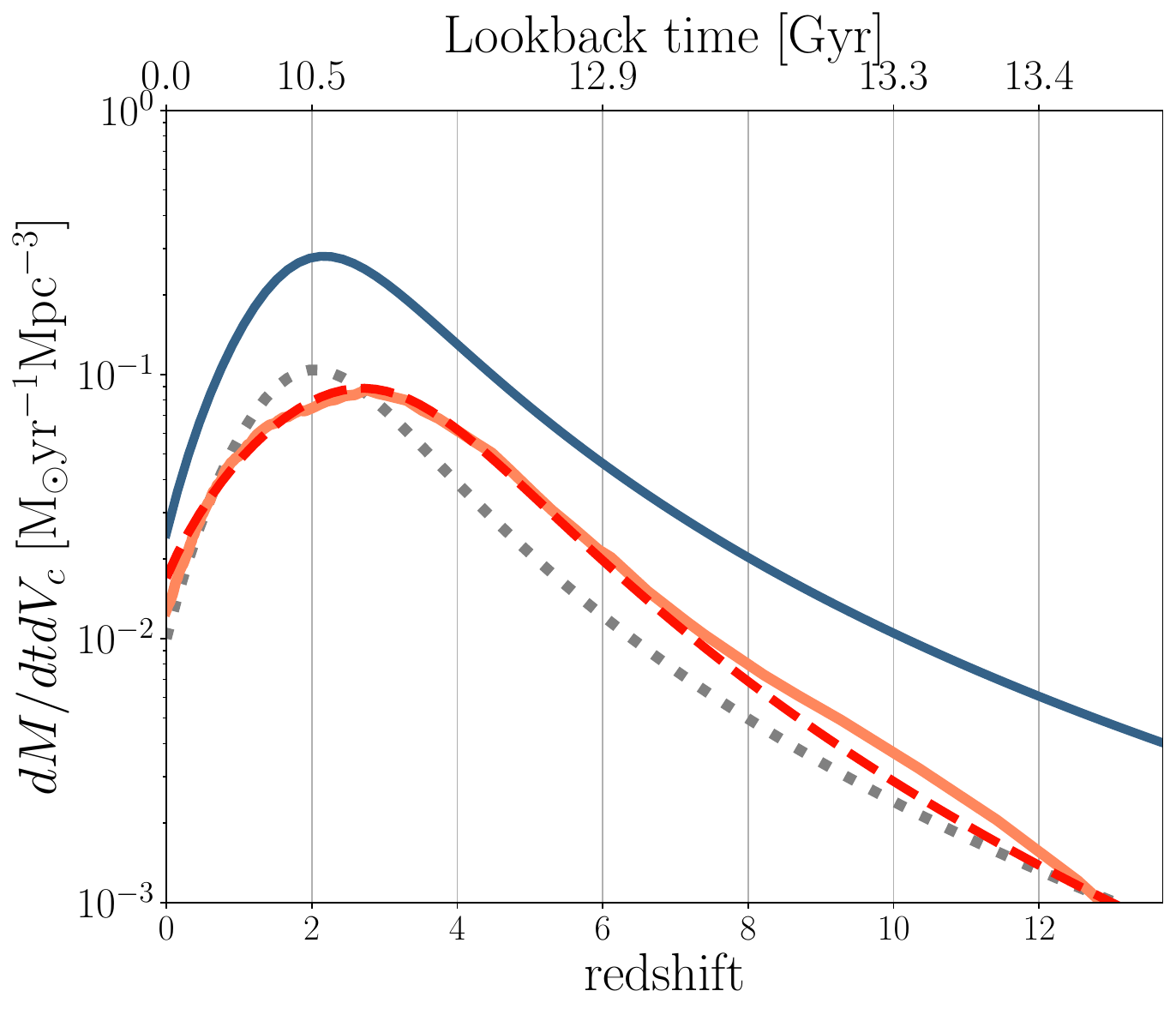}
\includegraphics[width=0.47\textwidth]{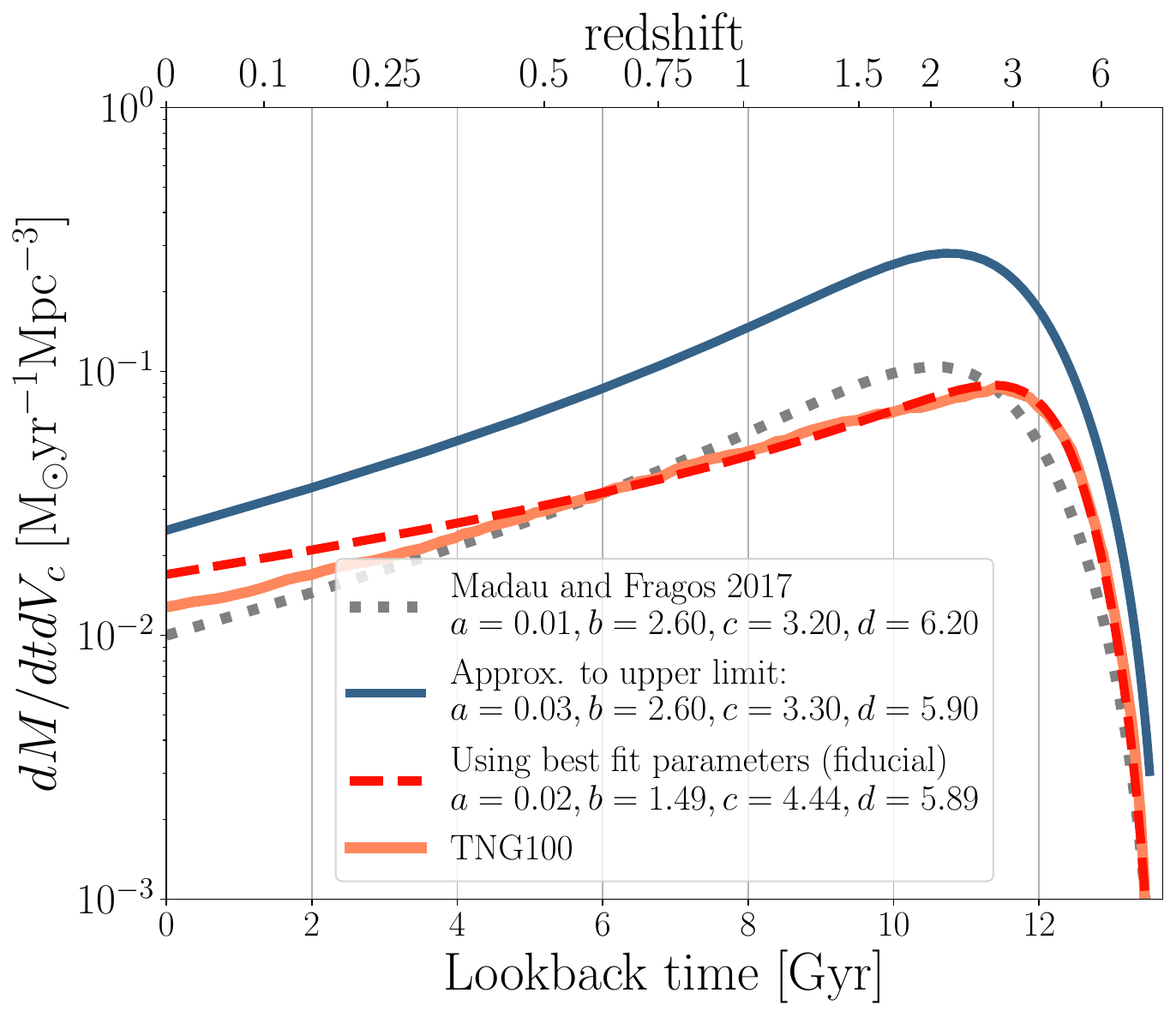}
\caption{Comparison of several overall star-formation rate densities, \SFRDz,  with redshift (left panel) and with lookback time (right panel). The solid orange and dashed red lines respectively show the star formation data from TNG100 and our corresponding fit adopting eq. \ref{eq: sfr1} (fiducial model). The dotted gray and solid blue lines are variations of eq. \ref{eq: sfr1} used to approximate the lower and upper edge of possible star-formation histories. The dotted gray line shows the model from \cite{Madau+2017}, while the solid blue line mimics the behaviour of the powerlaw-fit to the \texttt{SB: B18/C17} variations with a non-evolving low-mass end of the galaxy stellar mass function from \cite{Chruslinska+2021}.  \label{fig: SFR(z)}}
\end{figure*}

\begin{figure*}
\centering
\script{Plot_StableMass_distributions.py}
\includegraphics[width=0.8\textwidth]{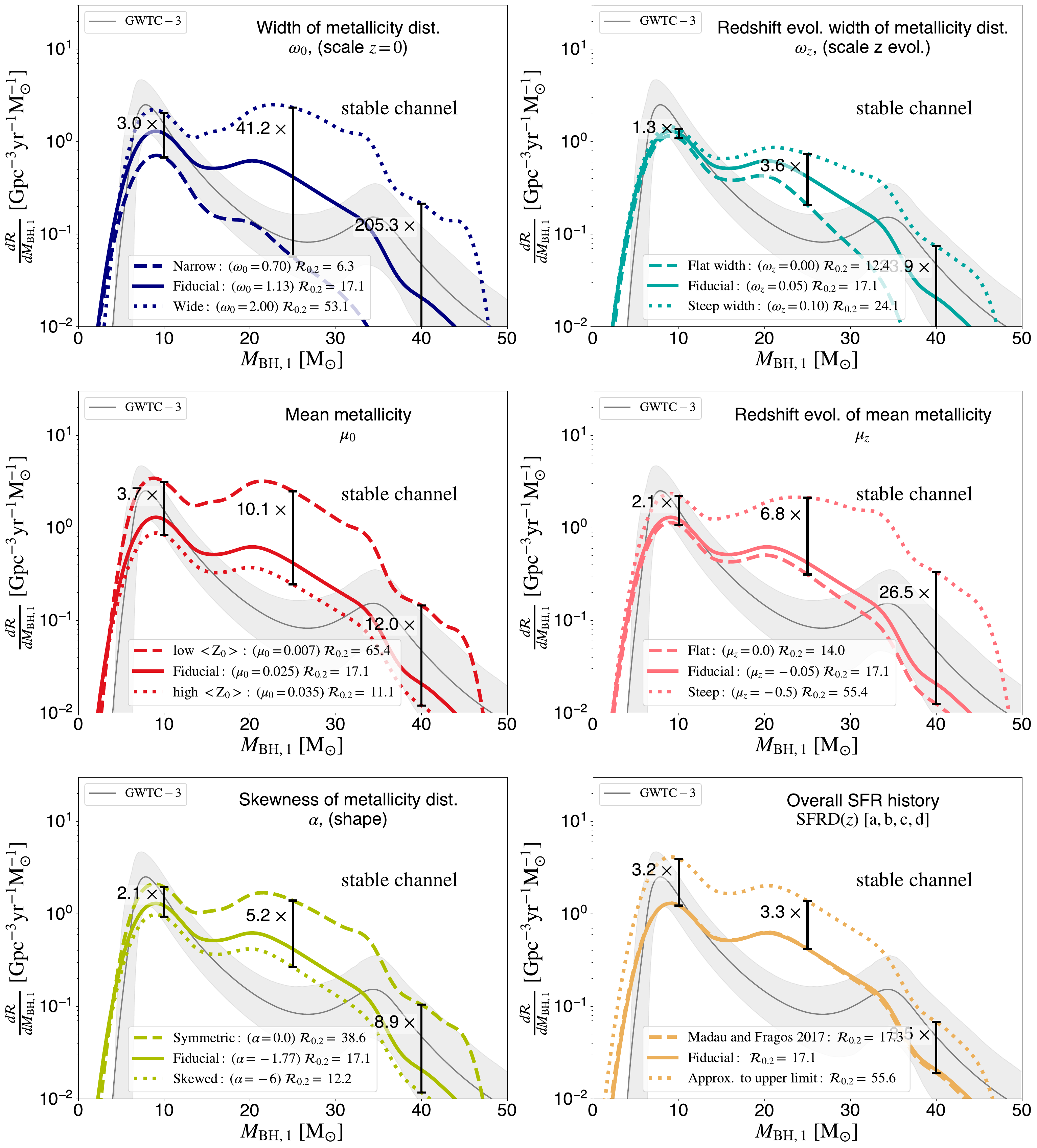}
\caption{The primary mass distribution of merging BBH systems from the stable mass transfer channel for several variations in \SFRDzZ. 
The first five panels show variations of the cosmic metallicity density distribution  \dpdZ, eq. \ref{eq: z log skew}, (parameters listed in the first three columns of Table \ref{tab: fit params}), where we vary one parameter at a time while keeping the rest fixed at their fiducial value. The bottom right panel shows variations in the magnitude of the star formation rate with redshift; i.e. \SFRDz. For the latter we vary the four fiducial parameters of \SFRDz simultaneously (last two columns of Table \ref{tab: fit params}). All panels are shown at a reference redshift of $z=0.2$, with the corresponding predicted BBH merger rate indicated in the legend. For reference, we show the power-law + peak model from \protect\cite{GWTC3_popPaper2021} in grey. We annotate the relative change in the rate at three reference masses: $10\Msun$, $25\Msun$ and $40\Msun$. 
  \label{fig: stable mass dists}}
\end{figure*}

We also consider two variations of the overall star-formation rate density, \SFRDz, where we keep the metallicity distribution \dpdZ fixed, but vary all four \SFRDz parameters at once (right two columns of Table \ref{tab: fit params}, and bottom of Table \ref{tab: fit variations}). 
We use Figure 11 from \cite{Chruslinska+2021} to determine approximate upper and lower bounds to the overall star-formation rate density. We choose \cite{Madau+2017} as an approximation of the lower limit. 
For the upper limit, we use the upper edge of models that adopt starbursts following \cite{Bisigello+2018} and \cite{Caputi+2017} (\texttt{SB: B18/C17}), combined with a non-evolving low-mass end of the galaxy stellar mass function \citep[shown as a thick brown line in Fig. 11 of ][and described in their table B1]{Chruslinska+2021}.
To approximate these models, we fit equation \ref{eq: sfr1} by eye to the broken power law description of this model as presented in appendix B1 of \cite{Chruslinska+2021}.
We show all \SFRDz variations in Figure~\ref{fig: SFR(z)}. \\

\begin{figure*}
\centering
\script{Plot_CEMass_distributions.py}
\includegraphics[width=0.8\textwidth]{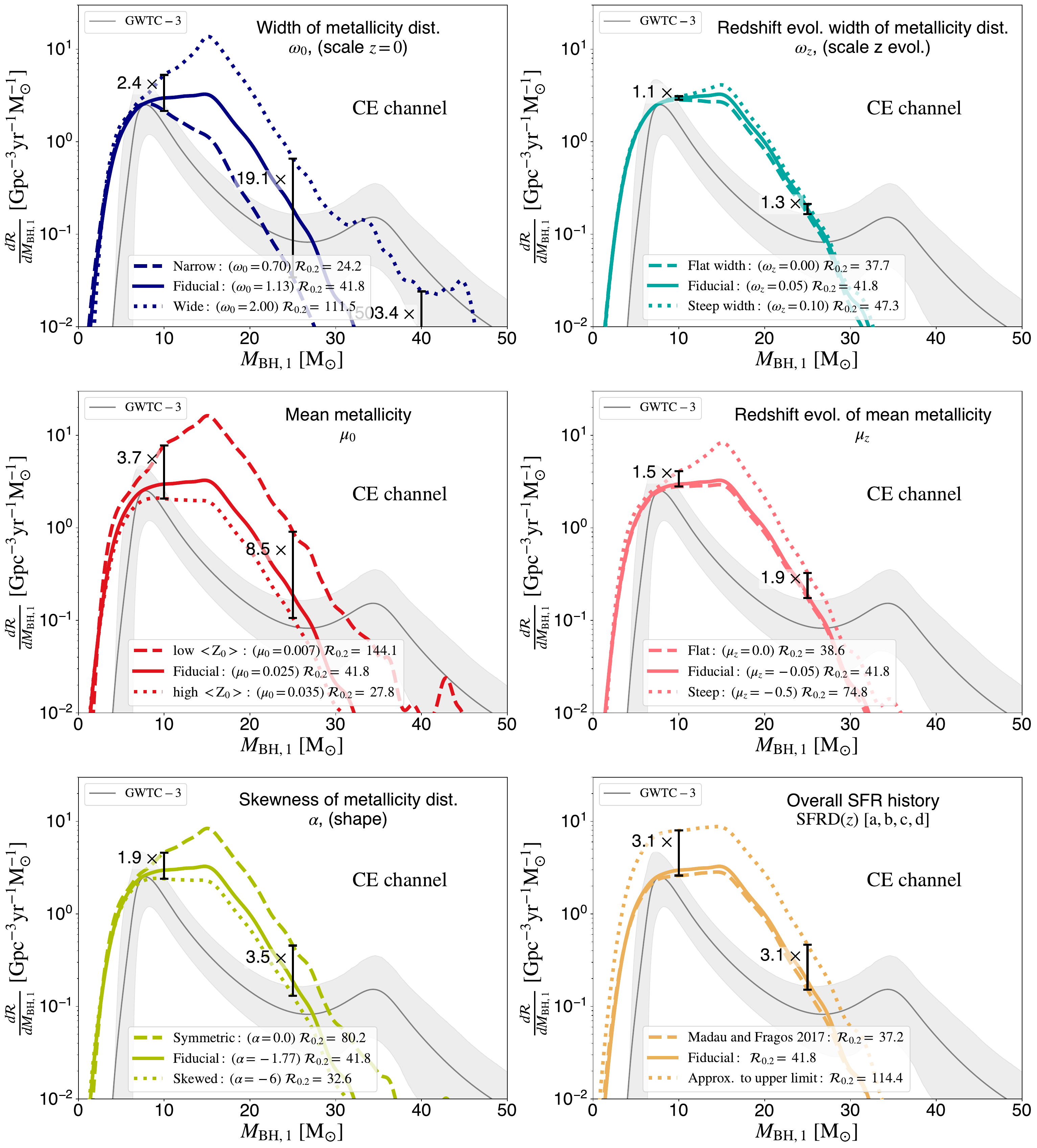}
\caption{Same as Figure \ref{fig: stable mass dists}, but for the Common Envelope channel. These figures show that the low mass end of the primary mass distribution is least affected by the adopted \SFRDzZ. Moreover, the \textit{location} of features in the mass distribution are robust against all explored variations.
   \label{fig: CE mass dists}
  }
\end{figure*}

\subsection{The effect of the \SFRDzZ on the primary masses of merging BBH}
To isolate the effect of the \SFRDzZ from the effects of different formation channels, we split the data from \cite{vanson+2022} between the stable mass transfer channel \citep[e.g.,][]{van-den-Heuvel+2017,Inayoshi2017,Bavera2021,Marchant2021,GallegosGarcia2021,vanson+2022}, and the `classical' common-envelope channel \citep[or CE channel, e.g.,\  ][]{Belczynski+2007,PostnovYungelson2014,Belczynski+2016Natur,vignaGomez+2018}. 
These channels are distinguished based on whether the binary system has experienced a common envelope phase (CE channel) or only stable mass transfer (stable channel in short from now on).\\

In Figures \ref{fig: stable mass dists} and \ref{fig: CE mass dists}, we show the resulting primary mass distribution of merging BBHs from the stable channel and CE channel respectively.
The primary (secondary) component refers to the more (less) massive component of merging BBHs. 
Each panel varies one aspect of the \SFRDzZ. In the first five panels of Figures \ref{fig: stable mass dists} and \ref{fig: CE mass dists}, we vary one of the  parameters that determine the shape of the probability density distribution of metallicities, while keeping all other values fixed at their fiducial values. In the last panel of  Figures \ref{fig: stable mass dists} and \ref{fig: CE mass dists}, we vary the shape of the overall star-formation rate densities, \SFRDz, to one of the variations shown in Figure \ref{fig: SFR(z)}, while keeping the probability density distribution of metallicities fixed.

The first thing we note is that the location of the features in the primary mass distribution are robust against variations in \SFRDzZ.
For the stable channel, two features are visible in all variations: a peak at $\Mbheen\approx9\Msun$ and a bump at $\Mbheen\approx22\Msun$. 
Two more features are visible in at the high mass end for almost all \SFRDzZ; a knee at $\Mbheen\approx35\Msun$ and another bump at $\Mbheen\approx45\Msun$. Although the locations of these features are constant, the features themselves can disappear for variations that suppress the rate of high mass BHs (e.g., dashed lines in the top panels of Fig. \ref{fig: stable mass dists}).  
Similarly, the CE channel displays a kink in the distribution at about $9\Msun$, and a peak at approximately $\Mbheen\approx17\Msun$ for all variations. The latter peak is the global peak of the mass distribution in almost all variations.  

The finding that the locations of features in the mass distribution do not change for different \SFRDzZ is consistent with earlier work. 
Recent work by \cite{chruslinska2022_review} showed that, when comparing two very different models of \SFRDzZ (their Figure 5), the location of the peaks remains the same, even though the normalisation between the two BBH merger rates is completely different. 
Furthermore, \cite{Broekgaarden+2021b} show the probability distribution of chirp masses for BBHs in their Fig. 4. Although features can disappear when the \SFRDzZ prohibits the formation of certain (typically higher) mass BHs, the \textit{location} of features remains the same.
This implies that the locations of features in the mass distribution of BBHs are determined by the formation channel and its underlying stellar and binary physics. The locations of features could therefore serve as sign posts of the underlying physics. \\

Second, we see that the low mass end of the primary mass distribution is relatively robust against variations in \SFRDzZ. 
To quantify this, we annotate the ratio between the maximum and minimum rate at three reference masses; \Mbheen = $10, 25$, and $40\Msun$.
At $\Mbheen=10\Msun$, we find that the rate changes by at most a factor of about 3.7 for the stable channel, and at most about a factor of 3.8 for the CE channel. 
On the other hand, the change in rate at $\Mbheen=40\Msun$ can be as high as a factor of about 200 and 150 for the stable and CE channels, respectively. 
The lowest mass BHs are least affected by the \SFRDzZ because they can be formed from all metallicities above $Z\gtrsim10^{-3}$ \citep[see e.g., Figures 7 and 13 from ][]{vanson+2022}.
The rate of star formation at metallicities above $\gtrsim 10^{-3}$ is observationally relatively well constrained for redshifts below $0.5$ (which comprises the past $5\Gyr$ of star formation). 
This is reflected in the top panel of Figure \ref{fig: low high Z fraction}: all models show that $10\%$ or less of the stellar mass was formed at a metallicity below $Z/10 \approx 0.0014$, or in other words, about $90\%$ or more of the stellar mass was formed at a metallicity above $Z/10$. Hence the lowest mass BHs derive from the least uncertain parts of the \SFRDzZ.
The low-mass end of the mass distribution of merging double compact objects will also provide a particularly powerful cosmological constraint in the era of third generation gravitational wave telescopes \citep{MariaEzquiaga2022}. Our finding that the low mass end is more robust against variations in \SFRDzZ supports this claim. 

Parameter variations that affect shape of \SFRDzZ at low redshift primarily change the normalisation of the mass distribution. This is the case for variations of the width of the cosmic metallicity density distribution at $z=0$ ($\omega_0$), the mean metallicity of the cosmic metallicity density distribution at $z=0$ ($\mu_0$), and the skewness of the cosmic metallicity density distribution ($\alpha$, left columns of Figures \ref{fig: stable mass dists} and \ref{fig: CE mass dists}).
To emphasise this point, we annotate the total BBH merger rate at redshift 0.2, $\mathcal{R}_{0.2}$, in the legends of Figures \ref{fig: stable mass dists} and \ref{fig: CE mass dists} (0.2 is the redshift where the observations are best constrained \citealt{GWTC3_popPaper2021}). 
Variations that increase the amount of star formation at low metallicity (i.e. for a low mean metallicity $\mu_0=0.007$ and a wide metallicity distribution $\omega_0 = 2.0$ ) increase the predicted BBH merger rate. This is consistent with other work that finds merging BBHs form more efficiently at low metallicities \citep[e.g.][]{BelczynskiVink2010, Stevenson+2017,Mapelli2017,Chruslinska2019_effectCO,Broekgaarden+2021b}.
A more skewed cosmic metallicity density distribution pushes the peak of the distribution to higher metallicities and thus forms more stars at high metallicity when compared to a symmetric distribution. Hence, the local rate of BBH mergers is lower for the skewed distribution ($\alpha = -6$) with respect to the symmetric variation ($\alpha = 0.0$).

Changing the overall star-formation rate density (\SFRDz, bottom right panels of Figures \ref{fig: stable mass dists} and \ref{fig: CE mass dists}) also affects the normalisation of the mass distribution, but has a smaller effect than the width and the mean of the cosmic metallicity density distribution at $z=0$ ($\omega_0$ and $\mu_0$). This underlines the importance of the amount of low-metallicity star formation \citep[e.g.,][]{chruslinska2022_review}, and is furthermore in line with findings from \cite{Tang+2020}.
As discussed in Section \ref{ss: reasonable var}, we use \cite{Madau+2017} and the solid blue line in Figure \ref{fig: SFR(z)} as an approximate lower and upper bound to the \SFRDz respectively. 
The overall cosmic star formation rate density from \cite{Madau+2017} is very similar to our fiducial model (Figure \ref{fig: SFR(z)}), and the differences between the resulting mass distributions are correspondingly small. Our approximation of the upper limit to the allowed \SFRDz leads to an overall increase of the BBH merger rate by a factor of about 3.

Parameters that change the evolution of the metallicity distribution \dpdZ with redshift, such as the redshift dependence of the with and mean; $\omega_z$ and $\mu_z$ (top right and centre right panels of Figures \ref{fig: stable mass dists} and \ref{fig: CE mass dists}) primarily affect the high mass end of the stable channel. 
We understand this as an effect of the different delay time distributions for both formation channels. Since both, $\omega_z$ and $\mu_z$ influence the amount of low metallicity stellar mass formed at high redshifts they will mostly affect systems with longer delay times. The stable channel has been shown to produce more high mass BHs with longer delay times when compared to the CE channel \citep{vanson+2022, Briel+2022}. 
Hence we find these variations affect the slope of the high mass end of the BBH mass distribution for the stable channel, while they have a relatively small impact on the CE channel.

\section{Discussion \& Summary \label{sec: summary}}
We present a flexible analytic expression for the metallicity-dependent cosmic star formation history, \SFRDzZ (equations \ref{eq: total sfrd}, \ref{eq: z log skew} and \ref{eq: sfr1}). An analytical expression allows for controlled experiments of the effect of \SFRDzZ on dependent values, such as the rate and mass distribution of merging BBHs. The model presented in this work adopts a skewed-lognormal for the distribution of metallicities at every redshift (\dpdZ). 

\paragraph{The model can capture the general behaviour of cosmological simulations, such as TNG100}
Our analytical expression for \SFRDzZ is composed of a cosmic metallicity density distribution that is determined by a mean, scale and skewness and their redshift dependence, as well as parameters governing the overall star-formation rate density. We fit our analytical expression for \SFRDzZ to the star-forming gas in the TNG100 simulation, and provide the best fit parameters in Table \ref{tab: fit params}. 
We show that our model captures the shape and general behaviour of the cosmological simulations well (Figure \ref{fig: fit SFRD}). 
Although our model is more broadly applicable than previous models, we acknowledge that it does not capture the \textit{complete} range of observationally-allowed variations in it's current form. Incorporating more complex functions for the redshift evolution of the metallicity could solve this issue, but this is left for future research.

\paragraph{The model allows for a controlled experiment on the effect of \SFRDzZ on the local distribution of merging BBH}
As an example, we use our model to calculate the local rate and mass distribution of the more massive components from merging BBHs (\Mbheen) in Figures \ref{fig: stable mass dists} and \ref{fig: CE mass dists}.
We systematically vary all five parameters that shape the cosmic metallicity density distribution, and explore two additional variations of the overall star-formation rate density \SFRDz.
Our main findings are as follows:
\begin{itemize}
    
    \item The locations of features in the distribution of primary BH masses are robust against variations in \SFRDzZ. The location of features in the mass distribution of BHs could thus be used as sign posts of their formation channel. 
    
    \item For all variations, the low mass end of the mass distribution is least influenced by changes in the \SFRDzZ.
    This is because the lowest mass BHs can be formed from all metallicities above $Z~\gtrsim~10^{-3}$, for which the star formation rate is relatively well constrained in the recent Universe. 
    This suggests that the lower end of the BH mass distribution (component masses of $\leq 15\Msun$) is potentially very powerful for constraining the physics of the formation channels, irrespective of the cosmic star formation rate uncertainties.
    
    \item The metallicity distribution of star formation at low redshift primarily impacts the normalisation of the BBH merger rate. Changing the overall star-formation rate density, \SFRDz also affects the rate, but to a lesser degree. This shows that low-metallicity star formation at low redshifts dominates the overall normalisation of the BBH merger rate. 
    
    \item Parameters that influence the redshift evolution of the mean and the width of the metallicity distribution affect the slope of the high mass end of the primary BH mass distribution for the stable channel. This reflects the longer delay times of the stable channel with respect to the CE channel. 
    
\end{itemize}

The flexibility of the model presented in this work can capture the large uncertainties that remain in the shape and normalisation of the metallicity-dependent cosmic star formation history. 
Our hope is that this expression will provide a useful starting point for making predictions and comparisons with observations.

\begin{acknowledgments}
The authors acknowledge partial financial support from the  National Science Foundation under Grant No. (NSF grant number 2009131  and PHY-1748958).”
, the Netherlands Organisation for Scientific Research (NWO) as part of the Vidi research program BinWaves with project number 639.042.728 and the European Union’s Horizon 2020 research and innovation program from the European Research Council (ERC, Grant agreement No. 715063). 
This research was supported in part by the National Science Foundation under Grant No. NSF PHY-1748958. 
\end{acknowledgments}

\section*{Software and Data}
All code associated to reproduce the data and plots in this paper is publicly available at \url{https://github.com/LiekeVanSon/SFRD_fit}.
The data used in this work is available on Zenodo under an open-source Creative Commons Attribution license at \dataset[10.5281/zenodo.7612755]{https://zenodo.org/record/7612755}.
All observationally constrained models of the \SFRDzZ from \cite{Chruslinska+2021} can be found online at: \url{https://ftp.science.ru.nl/astro/mchruslinska/Chruslinska_et_al_2021/}.

This research has made use of GW data provided by the Gravitational Wave Open Science Center (\url{https://www.gw-openscience.org/}), a service of LIGO Laboratory, the LIGO Scientific Collaboration and the Virgo Collaboration. 
Further software used in this work: Python \citep{PythonReferenceManual},  Astropy \citep{astropy:2013,astropy:2018} Matplotlib \citep{2007CSE.....9...90H},  {NumPy} \citep{2020NumPy-Array}, SciPy \citep{2020SciPy-NMeth}, \texttt{ipython$/$jupyter} \citep{2007CSE.....9c..21P, Kluyver2016jupyter},  Seaborn \citep{waskom2020seaborn}  and  {hdf5}   \citep{collette_python_hdf5_2019}.

\appendix

\section{Evaluating our fit; the squared residuals \label{app: residuals} }

\begin{figure}[h]
\centering
\script{Fit_model_to_sfrdzZ.py}
\includegraphics[width=0.49\textwidth]{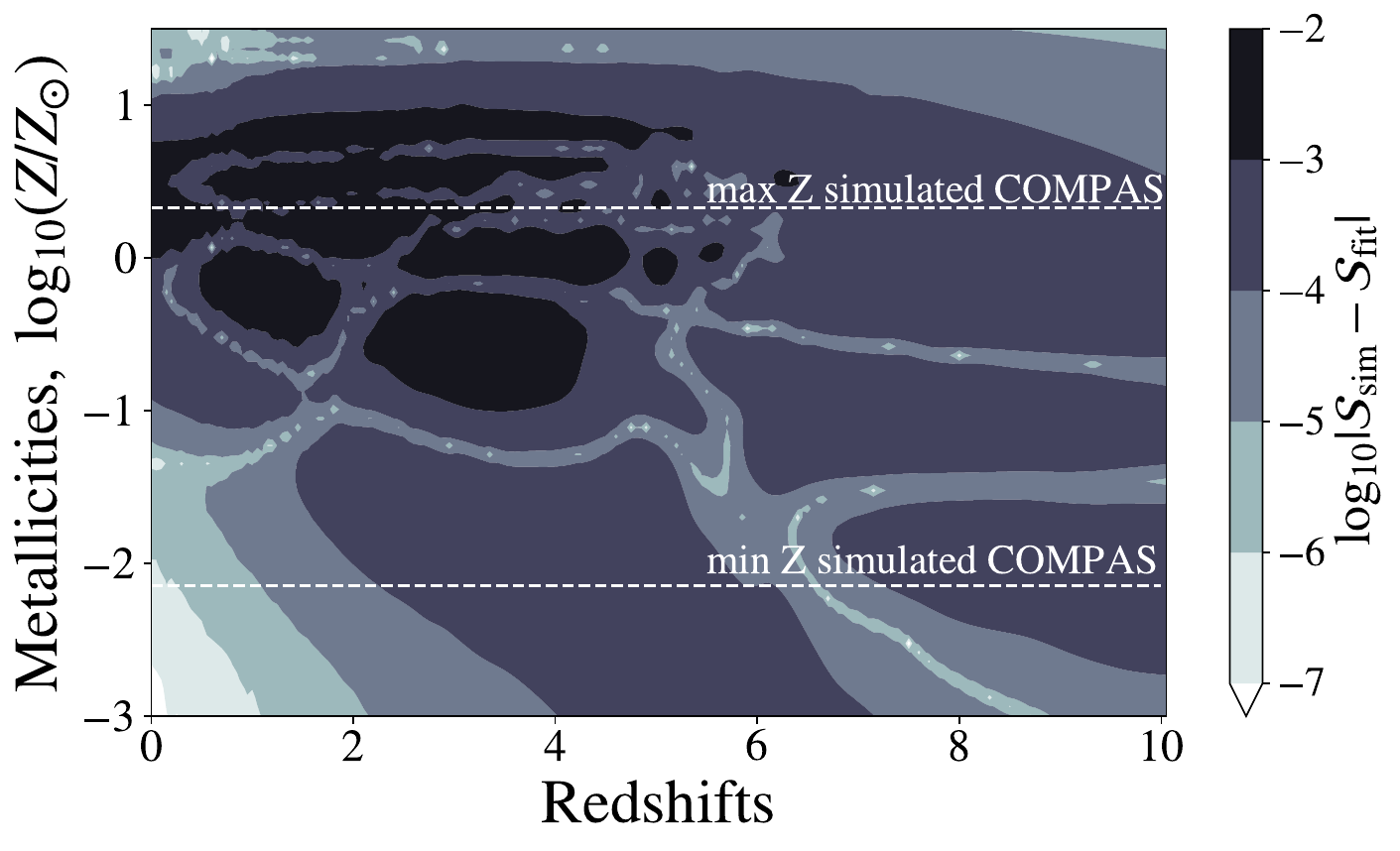}
\includegraphics[width=0.49\textwidth]{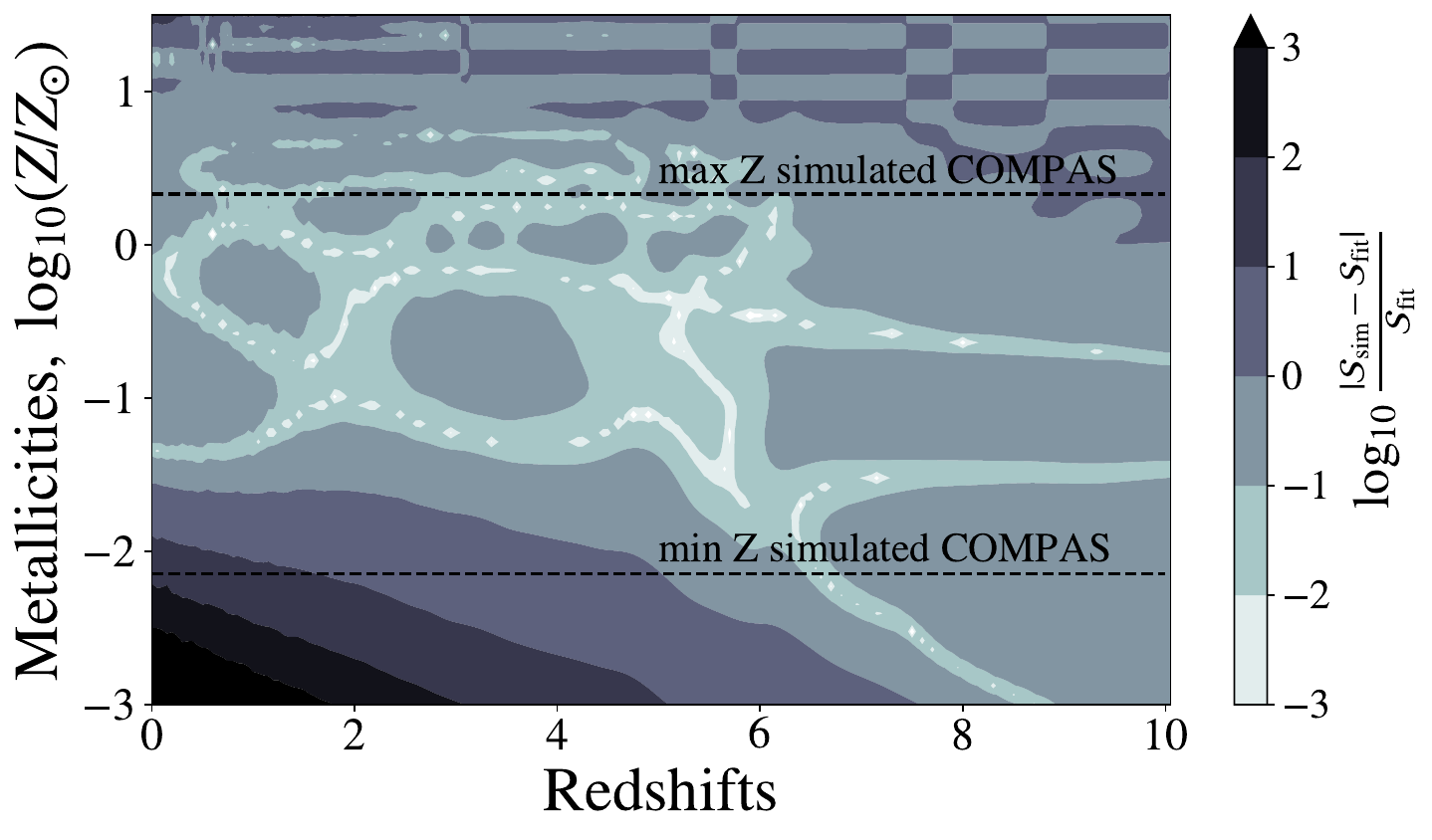}
\caption{$\log$ of the residuals (left), and $\log$ of the relative error (right) between the TNG100 data and our best-fitting model. We show the minimum and maximum metallicity used in \COMPAS simulations with dashed lines in each plot.  \label{fig: residuals}
 }
\end{figure}

In the left panel of Figure \ref{fig: residuals}, we show the $\log$ of the absolute residuals. The \textit{square} of the residuals is used in the cost function, equation \ref{eq: cost function}, to optimise our fit. We observe that the maximum residuals appear near the peak of star formation at high metallicities.
The log of the relative errors (defined as $\frac{\lvert \mathcal{S}{\rm sim} - \mathcal{S}{\rm fit} \rvert}{\mathcal{S}_{\rm fit}}$), is shown in the right-hand panel of Figure \ref{fig: residuals}. The relative errors generally exhibit an opposite trend with respect to the residuals. The relative errors are largest in regions of very low-metallicity star formation at low redshift. This occurs due to the very low star-formation rate in this regime (of the order $10^{-8}\Msun \yr^{-1}\Mpc^{-3}$ for the TNG simulations and $10^{-11}\Msun \yr^{-1}\Mpc^{-3}$ in our model fit).
Another regime where the relative error becomes large is at very high metallicities (about 10 times $Z_{\odot}$). This is because in this regime, the TNG data is very sparse and contains regions where the rate abruptly drops to zero. To avoid sharp features in the data, we use interpolated TNG data to produce the fit. 
We note that we chose to minimise the squared residuals (which is similar to minimising the mean squared error) in favour of minimising, for example, the relative error, to prevent overfitting such regions of very low star-formation rate. For the illustration purposes in this work, we are most interested in closely fitting the \SFRDzZ between the minimum ($10^{-4}$) and maximum ($0.03$) metallicities that can be simulated with \COMPAS. For applications that focus on extremely low ($<0.01 Z_{\odot}$) or extremely high ($\sim 10\times Z_{\odot}$) metallicity star formation, a different cost function would be more appropriate.

\section{Determining reasonable variations of the \SFRDzZ \label{app: reasonable var}}
\begin{figure}
\centering
\script{LowHighZ_fraction.py}
\includegraphics[width=0.75\textwidth]{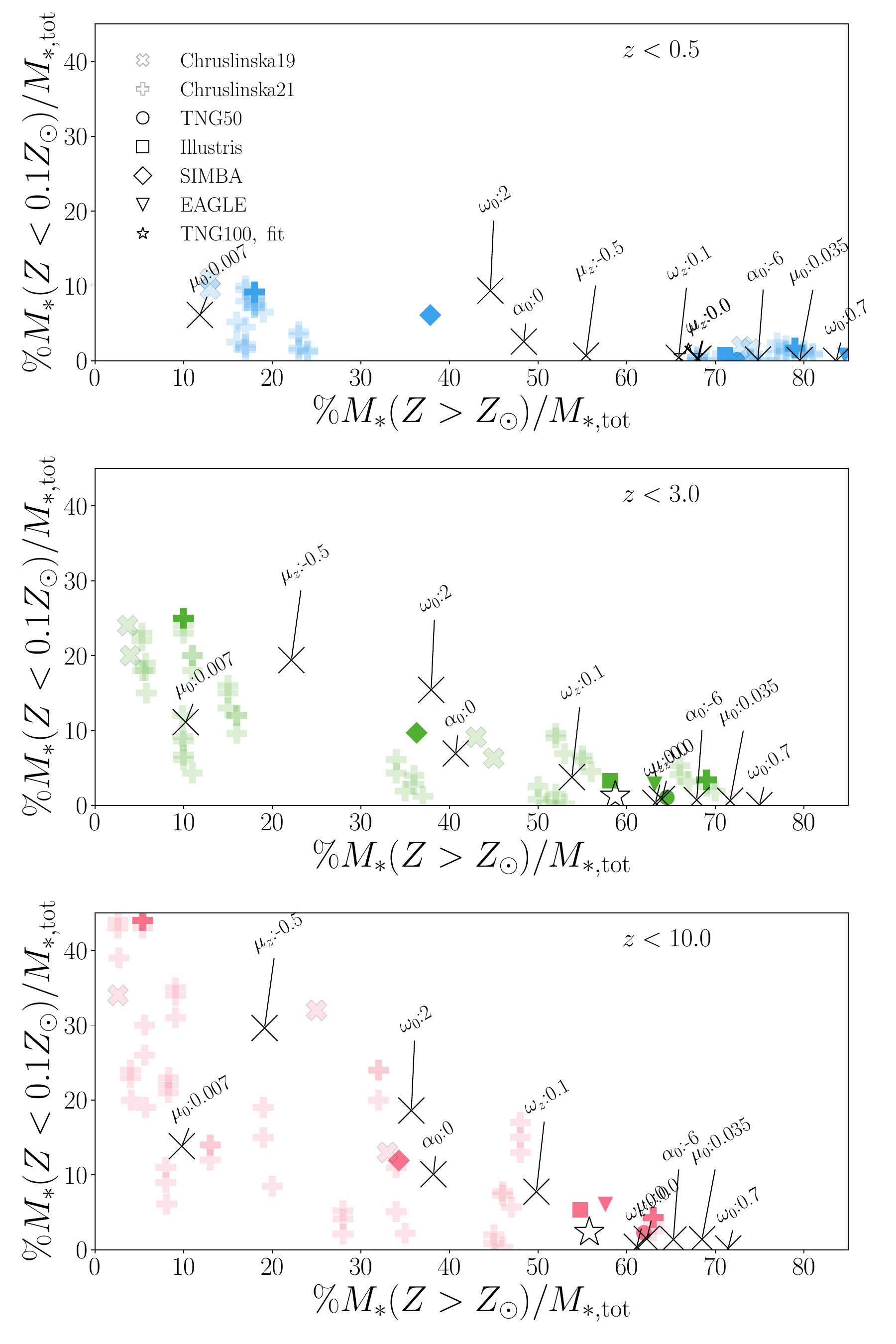}
\caption{Percentage of stellar mass formed at low metallicity ($Z < 0.1 \rm{Z_{\odot}}$) , versus high metallicity ($Z>\rm{Z_{\odot}}$) for all star formation below a certain threshold redshift: $z<0.5$ (top), $z < 3.0$ (middle) and $z<10$ (bottom). Data from observation-based variations are shown with semi-transparent thick crosses,  \citep[][]{Chruslinska2019_obs} and semi-transparent thick plus signs \citep[][]{Chruslinska+2021}, the low- and high-metallicity extremes are indicated with opaque symbols. For data from cosmological simulations, we follow \cite{Pakmor+2022} and show Illustris \citep[][squares]{Vogelsberger+2014}, Simba \citep[][diamonds]{Dave+2019}, EAGLE \citep[][triangles]{Schaye+2015}, TNG50 and TNG100 \citep[][filled and open circles respectively]{FirstResTNG_Springel2018}. Black thin crosses display variations of the cosmic metallicity density distribution that is part of our fiducial \SFRDzZ. The parameter that is varied with respect to the fiducial and its new value are annotated. This shows that our \SFRDzZ variations span the range of reasonable cosmic metallicity density distributions as determined by observation-based and cosmological simulations-based models.
\label{fig: low high Z fraction}
  }
\end{figure}

To determine reasonable variations of our fiducial model for \SFRDzZ, we compute the fraction of low and high metallicity stellar mass formed for redshifts below $z<0.5$, $z < 3.0$ and $z<10$. We show the results in Figure \ref{fig: low high Z fraction}, which is an adaptation of Fig. 2 in \cite{Pakmor+2022}, which in turn builds on Fig. 9 from \cite{Chruslinska2019_obs}.

\clearpage
\bibliography{main_bib}

\end{document}